\documentclass[aps,pre,twocolumn,groupedaddress,floatfix,showpacs]{revtex4}
\usepackage{amsfonts} \usepackage{amsmath}
\usepackage{float}

\usepackage{graphicx}

\newcommand{\B}{\mathbb{B}}

\renewcommand{\phi}{\varphi}

\DeclareMathOperator{\arcosh}{arcosh}
\DeclareMathOperator{\hw}{hw}

\begin{document}


\title{General properties and analytical approximations of photorefractive solitons}


\author{A. Geisler, F. Homann, H.-J. Schmidt}
\affiliation{Department of Physics, University of Osnabr\"uck,
  Barbarastrasse 7, D-46069 Osnabr\"uck}


\date{\today}
\begin{abstract} We investigate general properties of spatial $1$-dimensional
bright photorefractive solitons and suggest various analytical
approximations for the soliton profile and the half width, both
depending on an intensity parameter $r$.
\end{abstract}

\pacs{42.65.Tg, 05.45.Yv}


\maketitle

\section{Introduction}
\label{sec:introduction}

One-dimensional bright photorefractive solitons have been the subject
of numerous investigations by experimental and theoretical physicists \cite{Segev92,Segev94,Duree93}.
While experimentalists are primarily concerned with half width
measurements leading to so-called ``existence curves" \cite{Segev96},
there is also a theoretical interest in additional properties
of these solitary waves, such as the form and the asymptotic behavior
of the soliton profile with respect to large distances
or extreme values of its parameters. In view of the
lack of exact solutions of the relevant partial differential equation
there is hence a strong desire for proper approximations that suit the needs
of both groups of physicists alike.

To our knowledge, there is only one published approach to analytically approximating the soliton
profile, namely \cite{mm}. But this approach is not the only possible one:
In the present paper we will provide alternative approximations and discuss their respective virtues.
Further we will prove some general basic properties of photorefractive solitons which are independent of the chosen
approximation.
 
The starting point for our investigations is the theory developed by
Christodoulides and Carvalho \cite{cc} in which the profile $f(x)$ of bright
spatial photorefractive solitons is described by the 
following dimensionless differential equation (cf.~eq.(19) in
\cite{cc}):
\begin{equation}
 \label{1.1}
 f_{xx}  +  \beta F(f)  =  0,
\end{equation}
where
\begin{equation}
\label{1.2}
F(f)
=  2 f  \left( \frac{\ln(1 + r)}{r}  -  \frac{1}{1 + r f^2}  \right).
\end{equation}
Here $f$ is proportional to the electric field normalized to the
maximal value $1$, $r$ represents the ratio of intensity to dark
intensity of the beam and $x$ is the coordinate transversal
to the direction of the light beam. Since the factor $\beta$ can be
compensated by an appropriate scaling of the $x$-axis,
i.~e.~$x\to\sqrt{\beta}x$, we will choose $\beta=1$ throughout the
rest of the paper.
Thus $r>0$ is the only parameter the soliton profile $f(x)$ is depending
on. 

For the derivation of this nonlinear wave equation and the pertinent
simplifications we refer the reader to the original paper \cite{cc}.

Appropriate initial conditions for (\ref{1.1}) are
\begin{equation}
f(0)=1,\quad f_x(0)=0\quad .
\label{1.3}
\end{equation}
Equation (\ref{1.1}) is formally identical to a $1$-dimensional
equation of motion with a ``force function" $-F(f)$ and can be solved
analogously: One integrates (\ref{1.1}) once and derives an ``energy
conservation law"
\begin{equation}
 \frac{1}{2} f_x^2  +  V(f)  =  0\quad ,
\label{1.4}
\end{equation}
where the ``potential"
\begin{equation}
\label{1.5}
 V(f)
 =   f^2  \left(\frac{\ln(1 + r)}{r}  -
      \frac{\ln(1 + r f^2)}{r f^2} \right)
\end{equation}
has been introduced and the total ``energy" has been set to $0$ in
order to enforce the decay property $f(x)\rightarrow 0$ for
$|x|\rightarrow \infty$ for bright solitons. Separation of variables
yields the usual integral representation of the inverse function
$x(f)$:
\begin{equation}
\sqrt{2} x(f)=\pm \int_{1}^f \frac{d{\tt f}}{\sqrt{|V({\tt f})|}}.
\label{1.6}
\end{equation}
The half width $\hw(r)$ is defined as the length of the interval where
the intensity $f^2(x)$ exceeds half its maximal value, i.~e
\begin{equation}
\hw(r)\equiv 2 x \left( \frac{1}{\sqrt{2}} \right) =   \sqrt{2}
\int_{1/\sqrt{2}}^1 \frac{df}{\sqrt{|V(f)|}}
\quad .
\label{1.7}
\end{equation}
Although the integral (\ref{1.6}) cannot be solved in closed form, it
can be used to obtain numerical solutions of the soliton profile for
any given value of $r>0$.  One has to be careful because of the
(integrable) singularity of the integrand at $f=1$ of the form
$\frac{c_2}{\sqrt{1-f}}$, but most integration routines can deal with
such singularities.

However, for some purposes it is more convenient to work with closed
formulas for $f(x)$ or $\hw(r)$, albeit not exact ones, than with
numerical integrations.  Photorefractive soliton profiles and
existence curves have been measured over a range of six orders of
magnitude or $r$, cf.~\cite{Meng97, Kos98, Wesner01, WesnerDiss}.
Usually the experimental error margin for the measured values of
$\hw(r)$ is larger than the difference between an analytical and a
numerical approximation of $\hw(r)$. Similar remarks apply to the
soliton profile.  Hence for a comparison of experimental data with
theoretical predictions, the analytical approximation would be equally
good or even preferable, as long as it does not become too bulky.

\begin{figure}[H]
  \centering \includegraphics[width=\columnwidth, clip=true]{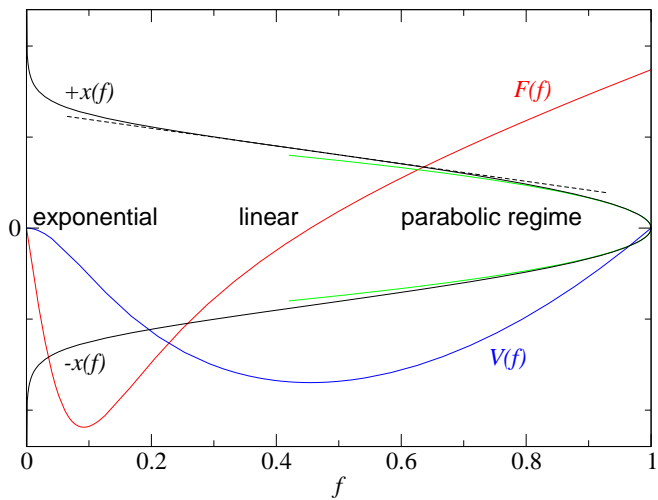}
  \caption{The typical form of functions $F(f)$ (\ref{1.2}), $V(f)$ (\ref{1.5})
  and $x(f)$  (\ref{1.6}) for a parameter value of $r=100$. Further, the
  different regimes are indicated according to the discussion in the introduction.}
  \label{fig:sqfig}
\end{figure}

Another aspect of the theory of optical solitons is the following:
Although (\ref{1.6}) cannot be solved in closed form, it is
nevertheless possible to exactly derive some characteristic properties
of the soliton which only depend on the differential equation.
This allows a semi-quantitative description of the soliton
amplitude $f(x)$. It starts at its maximum value $f(0)=1$
and decreases parabolically with the negative curvature
$f_{xx}(0)= -F(1)$ (parabolic regime). Then the curvature
approaches $0$ and the decrease of $f(x)$ is slower than parabolic.
The amplitude $f_w$ and the slope $f_{x}^{(w)}$ at the point of inflection,
i.~e.~where $f_{xx}$ vanishes, can be exactly determined via
(\ref{1.1}) and (\ref{1.2}). In the neighborhood of the
point of inflection the soliton profile is nearly linear (linear regime).
For $f<f_w$ the curvature  becomes positive and the curve  $f(x)$
is bent away from the $x$-axis. Finally, for $|x|\rightarrow \infty$,
the soliton amplitude decays exponentially (exponential regime),
where the  decay constant can be determined as a simple function of $r$, see
section \ref{gen}. This semi-quantitative discussion is illustrated
in figure \ref{fig:sqfig}.

Further the integral (\ref{1.6}) can be solved analytically for the
limits $r\rightarrow 0$ (giving the Kerr soliton) and $r\rightarrow
\infty$.

These partial analytical results, although being rather elementary,
are not easily found in the literature and thus appear worth while
mentioning in this article, see section \ref{gen}.

Apart from practical purposes it seems interesting that a power series
solution of (\ref{1.6}) can be obtained which allows, in principle, an
arbitrarily exact calculation of soliton profiles independent of the
intrinsic errors of numerical integration. However, the terms of this
series are of increasing complexity and we do not prove the series'
convergence. 

The article is organized as follows: In section \ref{basic} we resume
the partial analytical results mentioned above, including the
asymptotic solutions for $r\rightarrow 0$ and $r\rightarrow \infty$.
Section \ref{bounds} presents exact upper and lower bounds for the
half width $\hw(r)$ which are rather close, especially for $r>1$.
Section \ref{aa} contains an outline of the approximation devised by
Montemezzani and G\"unter \cite{mm} as well as two new analytical
approximations of $f(x)$ and 
the implied approximations of $\hw(r)$ which are of limited accuracy
but relatively simple. The first one, called $V$-approximation,
approximates the potential $V(f)$ by a cubic polynomial in $f^2$ such
that the integral (\ref{1.6}) can be done. The second one approximates
the integrand $1/\sqrt{|V(f)|}$ of (\ref{1.6}) by splitting off the
two poles at $f=0$ and $f=1$ and replacing the remaining function
$R(f)$ by the constant $R(1/2)$. It will be called
``$I$-approximation". Both methods give reasonable approximations of
$\hw(r)$ which suffice for practical purposes and can be considered as
alternatives to the approximations devised in Ref.~\cite{mm}.

In section \ref{sec:n-th-order-I} we complete the $I$-approximation by
a Taylor expansion of $R(f)$ about the center $f=1/2$ and show some
examples of approximate soliton profiles.  The number of terms of the
Taylor series which are needed to achieve a good approximation of
$f(x)$ increases with $r$. The lengthy but explicit expressions of the
general Taylor coefficients can be obtained via the Fa\`a di Bruno
formula and are given in the appendix.

It is obvious that our methods of approximation are not confined to
the special form of the photorefractive nonlinearity in (\ref{1.2})
but could also be applied to other nonlinear Schr\"odinger equations
or nonlinear oscillation problems.


\section{\label{basic} General properties and partial analytical results}

\subsection{\label{gen}General properties}
The basic equation (\ref{1.1}) is invariant under spatial reflections
$x\mapsto x_0-x$ and translations into $x$-direction. Hence any
solution $f(x)$ satisfying the (symmetric) initial conditions
(\ref{1.3}) is necessarily an even function of $x$, corresponding to
the $\pm$ sign in (\ref{1.6}). Since $V(f)<0$ for $0<f<1$, equation
(\ref{1.4}) shows that $f(x)$ is a strictly decreasing function for
$x>0$.

The Taylor expansion of $f(x)$ about the centre $x=0$ starts with
\begin{align}
\label{2.1}
f(x) & =  1  -  \frac{1}{2} V'(1)  x^2 + \ldots
 \nonumber \\ &= 1 -
\left(
\frac{\ln(1+r)}{r} -
\frac{1}{1+r}
\right) x^2 +\ldots
\end{align}
The corresponding parabola $f_q(x) = 1 - \frac{1}{2} {V'}(1) x^2$
represents a lower bound of $f(x)$ since it has the maximal negative
curvature of $f(x)$. Hence also its half width $\hw_q(r)$ will be a
lower bound of $\hw(r)$:
\begin{equation}
\label{2.2}
\hw_q(r)=\sqrt{\frac{2(2-\sqrt{2})r(1+r)}{(1+r)\ln(1+r)-r}} < \hw(r)
\quad .
\end{equation}

Moreover, $\hw_q$ yields the qualitatively correct behaviour for $r \ll
1$ and $r \gg 1$:

\begin{subequations}
\begin{gather}
  \label{2.3a}
  \hw_q(r) \approx 2 \sqrt{ \frac{2-\sqrt{2}}{r}} 
  \approx 1.53 \frac{1}{\sqrt{ r}} \quad (r \ll
  1)\\[3mm]
  \hw_q(r) \approx 2\sqrt{\left( 1- \frac{1}{\sqrt{2}} \right)\frac{r}{\ln r}}
  \approx 1.08 \sqrt{\frac{r}{ \ln r}} \quad
  (r \gg 1)
\end{gather}
\end{subequations}
This has to be compared with the asymptotic expressions for $\hw(r)$,
see below.

For $f\ll 1$ equation (\ref{1.1}) assumes the asymptotic form
\begin{equation}
\label{2.3}
f_{xx}+V''(0) f =  f_{xx}+ 2 f \left(
\frac{\ln(1+r)}{r}-1
\right)=0
\end{equation}
which has the solution
\begin{equation}
\label{2.4}
f(x)= C \exp\left(
-\sqrt{|V''(0)|}x
\right),\quad |x|\rightarrow\infty.
\end{equation}
Hence all solitons considered show an asymptotic exponential decay
for $|x|\rightarrow\infty$, the decay constant being a simple function
of $r$.


\subsection{Asymptotics for $r\rightarrow 0$}

\begin{figure}
  \includegraphics[width=\columnwidth]{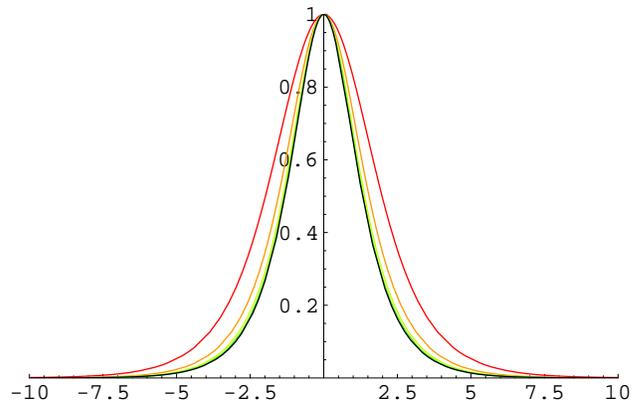}
\caption{\label{r0} Numerically determined soliton profiles for small values of $r$. If
  the $x$-axis is scaled with $\sqrt{r}$ they converge to the
  sech$(x)$ solution. The chosen values are $r=10^z$ with $z=0,
  -1/2,-1,-3/2$. }
\end{figure}

If $r \ll 1$, equation (\ref{1.1}) assumes the asymptotic form
\begin{equation}
{f''}   -    r f     (1  -    2 f^2)   =   0
\label{2.5}
\end{equation}
which is known from the cubic Schr\"odinger equation and has the
soliton solution
\begin{equation}
f_{0}(x)  =  \mbox{sech}(\sqrt{ r} x ) \quad.
\label{2.6}
\end{equation}
For the convergence of $f(x)$ towards the asymptotic form (\ref{2.6})
see figure \ref{r0}.

The corresponding half width is
\begin{equation}
\label{2.7}
\hw_{0}
 =  2 \frac{1}{ \sqrt{r} }  {\rm arcosh}\sqrt{2}
  \approx      \frac{ 1.76}{  \sqrt{r} }  \quad .
\end{equation}


\subsection{Asymptotics for $r\rightarrow \infty$}

\begin{figure}
  \includegraphics[width=\columnwidth]{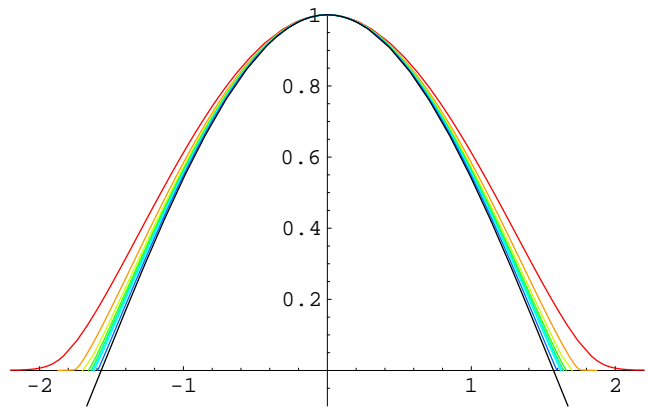}
\caption{\label{r00} Numerically determined soliton profiles for large values of $r$. If
  the $x$-axis is scaled with $\sqrt{\frac{2\ln r}{r}}$ they converge
  slowly to the $\cos(x)$ solution.  The chosen values are $r=10^z$
  with $z=3, 5, \ldots, 15, 30$. }
\end{figure}

If $r \gg 1$ and $f^2 \ln r \gg 1$, (\ref{1.1}) assumes the asymptotic
form of an oscillator equation
\begin{equation}    \label{2.8}
 {f''}  +  2   \frac{\ln r}{r}  f    =    0
\end{equation}
with the solution
\begin{equation}  \label{2.9}
f_\infty(x)  =
\begin{cases}
  \cos\left( \sqrt{2  \frac{\ln r}{r}}  x \right)  & |x|\le
  \frac{\pi}{2}\sqrt{\frac{r}{2 \ln r}}\\
  0 & |x|>  \frac{\pi}{2}\sqrt{\frac{r}{2 \ln r}}\; .
\end{cases}
\end{equation}
Here we have taken into account that for small $f$, $f^2 \ln r \gg 1$
as well as the approximation (\ref{2.8}) will be no longer valid and
the exponential decay will set in.  In (\ref{2.9}) this exponential
decay is approximated by setting $f_\infty(x)=0$ for $|x|> \frac{\pi}{2}\sqrt{\frac{r}{2 \ln r}}$.  The approach of the exact solution $f(x)$ to
(\ref{2.9}) for $r\rightarrow\infty$ is much slower than for the
analogous case $r\rightarrow 0$, see figure \ref{r00}. For a similar
result see \cite{chen}.

The corresponding half width is
\begin{equation}
\label{2.10}
\hw_\infty
 =  \sqrt{\frac{r}{2  \ln r}}   2 \arccos \frac{1}{\sqrt{2}}
 \approx  1.11 \sqrt{\frac{r}{\ln r}}.
\end{equation}


\section{Exact bounds for the half width}
\label{bounds}

We will utilise some properties of the (negative) force function
$F(f)$ introduced in (\ref{1.2}) which can be easily proven.  It has a
zero at
\begin{equation}
 \label{3.1}
f_0=\sqrt{\frac{r-\ln(1+r)}{r \ln(1+r)}} < \frac{1}{\sqrt{2}}
\end{equation}
which corresponds to the point of inflection $x(f_0)$ of the soliton
profile $f(x)$.  It follows that the half width is attained before the
point of inflection is reached, i.~e.
\begin{equation}
 \label{3.2}
 \hw(r)<x(f_0)\quad .
\end{equation}

The second derivative of $F(f)$ with respect to $f$ vanishes at
\begin{equation}
 \label{3.3}
f_w=\sqrt{\frac{3}{r}}\quad .
\end{equation}
Some simple calculations then show that $F$ is a convex function
within the physical domain $f \in [0, 1]$ if $r < 3$ and a concave
function within the domain $f \in [f_0, 1]$ if 
$r>r_0 \approx 9.3467$.
Here $r_0$ is the solution of $f_0(r)=f_w(r)$.  In both cases $F$ can
be bounded by affine functions of the form $a(f)=m (f-1)+F(1)$.  Note
that an affine force function of this form would lead to harmonic
oscillations $f(x)$ and a corresponding half width
\begin{equation}
\label{3.4}
\hw_m
 =
  \frac{2}{\sqrt{m}}  \arccos \left( 1  -  \frac{m}{F(1)} \left( 1 -
      \frac{1}{\sqrt{2}} \right) \right) 
\quad .
\end{equation}
Now assume an inequality between two force functions $F_1 < F_2$
within some domain.  By integration we conclude $|V_1| < |V_2|$ and, using
(\ref{1.6}), the reverse inequality $x_2(f)< x_1(f)$ for the positive
branch $x_i(f)> 0,\; i=1,2$.  Hence also $\hw_2 < \hw_1$ if the half
width is assumed within the domain under consideration.

By applying these arguments to our particular cases we obtain
\begin{equation}
\label{3.5}
 s(f)  >  F(f)  >  t(f)  \qquad \forall f \in (0, 1)   \qquad  (r < 3)
\end{equation}
where $t=t(f)$ is the tangent through the point $(1, F(1))$,
\begin{subequations}
\begin{gather}
\label{3.6}
  t(f)  = m_t  (f - 1)  +  F(1)  ,\;\text{with}\\[2mm]
  m_t \equiv  \left.  \frac{\partial F}{\partial f} \right|_{f = 1}=
  F(1)+\frac{4r}{(1+r)^2}, \label{3.6b}
\end{gather}
\end{subequations}
and $s(f)$ is the secant through $(1, F(1))$ and $(f_0, F(f_0))$,
\begin{equation}
\label{3.7}
  s(f)
  =  m_s   (f - 1)  +  F(1),  \qquad   m_s \equiv  \frac{F(1)}{1 - f_0}  .
\end{equation}
Consequently,
\begin{equation}
\label{3.8}
\hw_{m_s}   <   \hw  <  \hw_{m_t}  \qquad (r < 3)
\quad .
\end{equation}

In the case of $F$ being concave the inequalities (\ref{3.5}) and
(\ref{3.8}) are just inverted and we obtain
\begin{equation}
\label{3.9}
\hw_{m_t}   <   \hw  <  \hw_{m_s}  \qquad (r > r_0\approx 9.3467)
\end{equation}

\begin{figure}
  \includegraphics[width=\columnwidth]{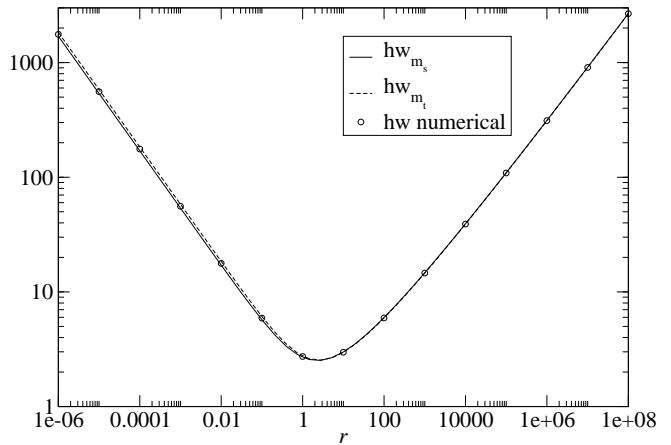}
\caption{\label{hws_hwt} Exact bounds $\hw_{m_s}$, $\hw_{m_t}$ for the
  half width according to \eqref{3.4}, \eqref{3.6b}, \eqref{3.7} and
  the numerically determined half width as functions of $r$.}
\end{figure}

Fig. \ref{hws_hwt} confirms in a double-logarithmic plot that the
numerically determined half width lies between $\hw_{m_s}$ and $\hw_{m_t}$.
For $r>10$ the two bounds almost coincide.

\section{Analytical approximations of the soliton profile}
\label{aa}

As explained in the introduction it would be desirable to have
analytical approximations of the soliton profile $f(x)$ and half width
$\hw (r)$ which are not too complex in form and yet give qualitatively
correct results for a large range of values of $r>0$. In this section
we will present the three approximations mentioned in the introduction.


\subsection{P-approximation, cf.~\cite{mm}}
We will call the approximation of the soliton profile $f(x)$ due to
G.~Montemezzani and P.~G\"unter ``P-approximation". It will suffice to
briefly sketch it and to refer the reader for more details to \cite{mm}.

The key idea is to expand the inverse soliton profile $1/f(x)$ into an even power series
$P(x)=\sum_{n=0}^\infty a_{2n} x^{2n}$.
Inserting this ansatz into (\ref{1.1}) allows the determination of arbitrary
coefficients $a_{2n}$ by means of recursion relations.
Hence this method yields, in principle, arbitrary precise approximations of $f(x)$.

However, there are -- to our opinion -- some minor disadvantages of the P-approximation
which motivate the development of alternative approximations:
\begin{itemize}
\item
For concrete approximations $P(x)$ has to be replaced by a polynomial of degree, say, $2n$.
In this case the exponential decay of $f(x)$ is not properly reproduced.
\item
The half-width $\hw(r)$ can be given by an explicit expression only for $n\le 4$,
in simple form even only for $n=2$, see \cite{mm}.
\item
We do not see any possibility to extend the P-approximation to the case of dark
solitons, in contrast to the according claim in \cite{mm}.
\end{itemize}


\subsection{$V$-approximation}

The $V$-approximation is an approximation of $V(f)$ in the
neighbourhood of $f=1$ which reproduces the zeros of $V$ (the double
zero $f=0$ and the simple zero $f=1$):

Introducing the abbreviation
\begin{equation}
\phi(u)
\equiv  \frac{\ln(1 + u)}{u} \quad ,
\label{4.1}
\end{equation}
the potential $V$ can be written as
\begin{equation}
\label{4.2}
 V(f)  =   f^2  \left( \phi(r)  -  \phi(r f^2) \right)   \quad .
\end{equation}
A Taylor series of $\phi$ with the centre $f=1$ including terms of
second order $(1 - f)^2$ yields the approximation
\begin{equation}
V_0(f)
  =   - \frac{1}{8} \alpha  f^2   \left(f^2 - 1\right)  \left( f^2 -
    f_1^2 \right) 
\label{4.3}
\end{equation}
where
\begin{equation}
 f_1^2
\equiv   \frac{-5 r^2 - 4 r  + 4 (1 + r)^2 \ln(1 + r)}
{- 3 r^2 - 2 r + 2 (1 + r)^2 \ln(1 + r)}
\label{4.4}
\end{equation}
and
\begin{equation}
\label{4.5}
 \alpha    \equiv   4     \frac{- 3r^2  -  2r  +  2(1 + r)^2   \ln(1 +
   r) }{r  (1 + r)^2} \quad . 
\end{equation}

By inserting the approximated potential $V_0$ into equation
(\ref{1.6}) and solving the integral we obtain the approximated
soliton intensity
\begin{equation}
\label{4.6}
f_V^2(x)  = \frac{2f_1^2}{f_1^2 + \left(f_1^2 -1\right)\cosh\left(\sqrt{ \alpha }    f_1  x \right)}
\end{equation}
and the half width
\begin{equation}
 \hw_V(r)  =  2 \frac{1}{\sqrt{\alpha } f_1}  \arcosh\frac{ 3  f_1^2
   -  1}{f_1^2 - 1} \quad . 
\label{4.7}
\end{equation}

Due to the choice of the centre $f = 1$ in the Taylor approximation
the soliton profile is well approximated in the neighbourhood of the
maximum $f = 1$ for all $r>0$. In fact, plots of $f(x)$ for different
$r$ show a good agreement of $f_V$ and $f$ if $\frac{1}{\sqrt{2}} \le
f \le 1$ for arbitrary $r$. This can be explained by a comparison with
the $r\rightarrow 0$ approximation $f_0(x)= \mbox{sech}(\sqrt{r} x)$
which yields the result
\begin{equation}
\label{4.8}
\frac{f_V(x)}{f_0(x)}=  1+ \frac{2}{3} x^2 r^2  +{\cal O}(x^2 r^3).
\end{equation}

Therefore we expect to find a good approximation of the soliton's half
width for all $r$.  Indeed, if $r \ll 1$, the result of the
$r\rightarrow 0$ approximation (\ref{2.7}) of the half width is
reproduced exactly:
\begin{equation}
\label{4.9}
\hw_V(r)
 =  \frac{2}{ \sqrt{r} }  \left(\arcosh\sqrt{2} + {\cal O}(r) \right)
 =  \hw_0(r)+ {\cal O}(\sqrt{r}).
\end{equation}

For large values of $r$, we find
\begin{align}
  \hw_V(r) & \approx \sqrt{\frac{r}{\ln(r) }}
  \frac{\ln(5  +  2 \sqrt{6})}{2}   \nonumber \\
  & \approx 1.1462 \sqrt{\frac{r}{ \ln(r) }} \qquad (r \gg 1) ,
  \label{4.10} 
\end{align}
which does not reproduce the result (\ref{2.10}) exactly, but yields a
good approximation.

On the other hand, the soliton profile $f(x)$ in the region $f \ll 1$
is only reproduced if $r \ll 1$, thus the $V$-approximation is not
suited to analyse the exponential decrease.

\subsection{$I$-approximation}
\newcommand{\zeroeth}{$0^\text{th}$}

According to \eqref{1.6} the soliton's shape is not directly governed
by the potential $V$ but by the integrand
\begin{equation}
  \label{eq:2}
  I(f)\equiv\frac{1}{\sqrt{-V(f)} }.
\end{equation}
Hence, an approximation of the integrand rather than the potential
itself might be a good starting point for approximating the soliton,
too.  Taking into account the poles at $f=1$ and $f=0$ it is a natural
idea to split the integrand up into three distinct parts as
\begin{equation}
  \label{eq:3}
  \frac{1}{\sqrt{- V(f)}} = \frac{c_1}{f} + \frac{c_2}{\sqrt{1-f}} + R(f),
\end{equation}
where a new function $R(f)$ and the two constants $c_1$ and $c_2$ have
been introduced.  The constants can be determined by
\begin{align}
  c_1 & \equiv \lim_{f \to 0} \frac{f}{\sqrt{- V(f)}}
  =   \frac{1}{\sqrt{- \frac{1}{2} {V''}(0)}} \nonumber \\[2mm]
  & = \left[ \left(1 - \frac{\ln(1 + r)}{r} \right) \right]^{-
    \frac{1}{2}}
\label{c1}
\end{align}
and
\begin{align}
  c_2 & \equiv \lim_{f \to 1} \frac{\sqrt{1 - f}}{\sqrt{- V(f)}}
  =   \frac{1}{\sqrt{{V'}(1)}} \nonumber \\[2mm]
  & = \left[ 2 \left( \frac{\ln(1 + r)}{r} - \frac{1}{1 + r} \right)
  \right]^{- \frac{1}{2}}.
\label{c2}
\end{align}
While the integrand's behaviour at the poles is correctly covered by
the first two summands of \eqref{eq:3} the function $R(f)$ dominates
the integrand in the region between the poles.  A good approximation
of the integrand can now be achieved by expanding the function $R(f)$
into a Taylor series. Due to its construction this fits the
exact soliton best for $f\approx 1$ and $f \ll 1$. 

\subsubsection{\zeroeth\/ order $I$-approximation}
\label{sec:0th-order-i}

Since $R(f)$ does not vary too much over the interval $0<f<1$ even a
\zeroeth\/ order approximation yields good results. $R(f)$ will be expanded
around $f=1/2$. Note that this choice is rather arbitrary but seems
reasonable.  Up to order 0 the $I$-approximation gives
\begin{equation}
  \label{eq:1}
  I(f)\approx \frac{c_1}{f} + \frac{c_2}{\sqrt{1-f}} + c_3,
\end{equation}
with
\begin{multline}
\label{c3}
c_3 \equiv
R \left( \frac{1}{2} \right)\\
= \frac{2\sqrt{r}}{\sqrt{4 \ln(4+r)-\ln(1+r)-8\ln 2}} -2
c_1-\sqrt{2}c_2 \;.
\end{multline}

By a simple integration one obtains
\begin{equation}
\label{xp(f)}
  - c_1 \ln f   +  2 c_2 \sqrt{1 - f}   +  c_3 (1 - f)    =
 \sqrt{2}  x,
\end{equation}
and the corresponding half width
\begin{multline}
  \hw_I(r) = \sqrt{2} \left( c_1(r) \ln \sqrt{2}
    +  2  \sqrt{1 - \frac{1}{\sqrt{2}}}\;  c_2(r) \right. \\
  + \left. \left(1 - \frac{1}{\sqrt{2}}\right) c_3(r) \right).
\label{hwp}
\end{multline}

For small $r$ a series expansion with respect to $r$ yields
\begin{equation}
 \hw_I(r)
 \approx   \frac{1.82}{\sqrt{ r }} \qquad (r \ll 1).
\end{equation}
Although the result of the low amplitude approximation (\ref{2.7}) is
not reproduced exactly, this is in fact very close to it.

For large values of $r$ we find
\begin{align}
  \hw_I(r) & \approx \sqrt{\frac{r}{2 \ln r}}
  2 \left[ \sqrt{2 - \sqrt{2}} \;+ \right. \nonumber \\[1ex]
  & \quad \left. \left(1 - \frac{1}{\sqrt{2}} \right)
    \left(\frac{2}{\sqrt{3}} - 1 \right) \right]
  \\[1ex]
  & \approx 1.1465 \sqrt{\frac{r}{\ln(r) }} \qquad (r \gg 1),
\end{align}
which matches the corresponding value of the
V-approximation up to three decimal places.

Finally, for $f \ll 1$, from eq.~(\ref{xp(f)}) one easily derives the
exponential decrease of the soliton:
\begin{equation}
f(x) \approx \exp\biggl( { \frac{2 c_2 + c_3 - \sqrt{2} x}{c_1}} \biggr) \qquad (f \ll 1).
\end{equation}

\begin{figure}[htbp]
  \includegraphics[width=\columnwidth, clip=true]{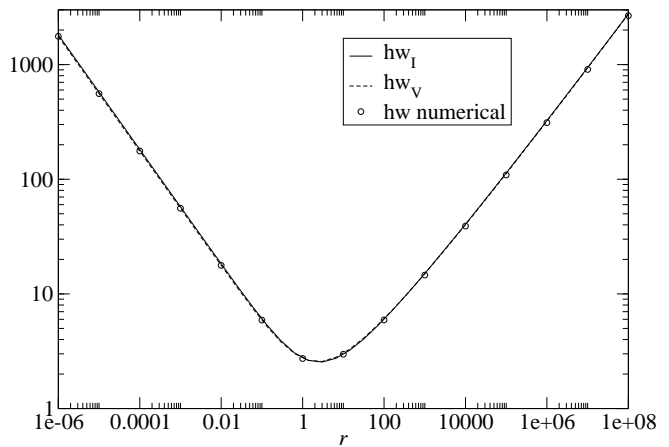}
\caption{\label{hwz_hwp} Approximated half widths $\hw_V$, $\hw_I$
  according to \eqref{4.7}, \eqref{hwp}  and the numerically
  determined half width as functions
  of $r$ -- almost indistinguishable.}
\end{figure}

Fig. \ref{hwz_hwp} shows that the approximated half width $\hw_V$ and
$\hw_I$ are in good agreement with the numerically determined
half width.

\begin{figure}[htbp]
  \centering \includegraphics[width=\columnwidth,clip=true]{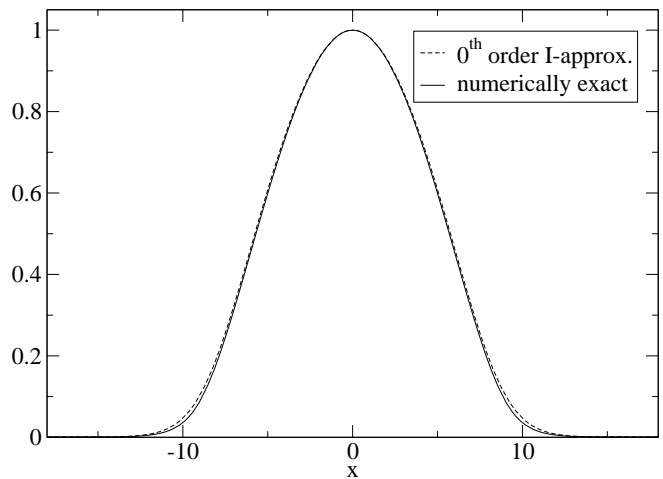}
  \caption{\zeroeth\/ order $I$-approximation and numerically exact soliton, $r=100$.}
  \label{fig:zerofig}
\end{figure}

The shape of the soliton is very well fitted by the \zeroeth\/ order
$I$-approximation as well, see Fig.~\ref{fig:zerofig}

\subsubsection{Higher order $I$-approximations}
\label{sec:n-th-order-I}

\begin{figure}[htpb]
  \includegraphics[width=\columnwidth, clip=true]{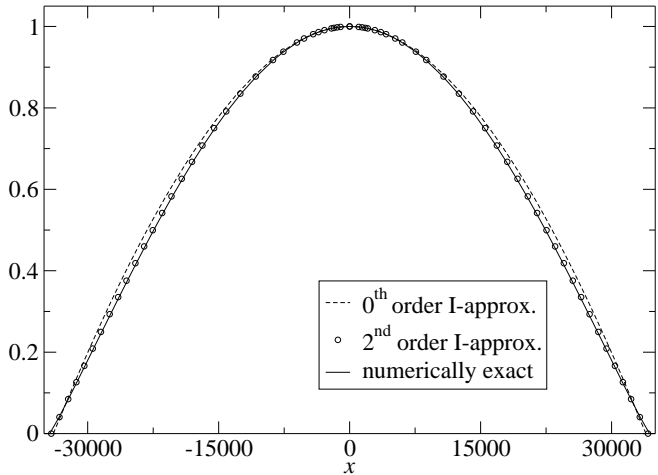}
  \caption{\zeroeth\/ and $2^\text{nd}$\/ order $I$-approximation
    together with the numerically exact soliton for $r=10^{10}$.}
  \label{fig:r1E10}
\end{figure}

Although for practical purposes the \zeroeth\/ order $I$-approximation gives
a sufficiently accurate approximation of the half width, the shape of
the soliton has still room for improvement, especially for large $r$.
The necessary enhancement of the $I$-approximation can easily be
achieved by taking higher order Taylor coefficients into account.
Fig.~\ref{fig:r1E10} shows that even for very
large $r$ a satisfactory approximation can be achieved with a second
order $I$-approximation.  If still higher order approximations are
needed the necessary Taylor coefficients are given in
Appendix \ref{sec:taylor}.

\section{Summary}
In this paper we considered the profile $f(x)$ of
spatial one-dimensional bright photorefractive solitons,
depending on an intensity parameter $r>0$,
which is only given by means of an integral (\ref{1.6})
but not in closed form. \\

We presented partial analytical results which allow a semi-quantitative discussion
of the profile and  studied the closed form solutions for the limit cases
$r\rightarrow 0$ and  $r\rightarrow \infty$. We also provided exact
bounds of the half width curve $\hw(r)$.\\

Moreover, we devised several analytical approximations of the
soliton profile and the half width which are relatively simple in form,
but in excellent agreement with the numerical results.
These approximations would thus suffice for the practical purpose
of comparing experimental data with theoretical descriptions.
If an arbitrary high accuracy of the approximation is desired
one has to resort to the Taylor series (\ref{eq:12}).
Altogether, we thus consider the problem of evaluating
the soliton profile (\ref{1.6})
as essentially solved.\\

We expect that our methods can be used as a basis to analyse more complex situations
such as photorefractive solitons under influence of diffusion and also be transferred
to the problem of dark and grey solitons.


\section*{Acknowledgement}
We would like to thank M.\ Wesner and H.-W.\ Sch\"urmann for
  critically reading the manuscript, stimulating discussions and
  helpful suggestions.
\appendix

\section{Taylor coefficients of $R(f)$}
\label{sec:taylor}

In order to give the explicit expression of the Taylor coefficients of
the integrand $I(f)$ and the rest function $R(f)$ we write them as the
composition of auxiliary functions that can be handled much better. Let
\begin{gather}
  g_1(x)=\ln(1 + r x^2),
  g_2(x)= x^2 \ln(1+r),\\
  g(x)= g_1(x)-g_2(x), \;\text{and} \\
  h(x)=\frac{1}{\sqrt{x} }.
\end{gather}
Then the integrand $I$ can be rewritten as
\begin{equation}
  \label{eq:4}
    I(f)= \sqrt{r} \cdot [h\circ g](f)
\end{equation}
The $n$th derivative of all the constituents of $I(f)$ can easily be
calculated. For $n\geq 1$ they read
\begin{gather}
g_2^{(1)}(x)= 2 x \ln(1+r), \quad g_2^{(2)}(x)=2 \ln(1+r),\\
n>2:\, g_2^{(n)}(x)=0, \\
g^{(n)}(x)=g_1^{(n)}(x) - g_2^{(n)}(x),\\
h^{(n)}(x)= 2\cdot \left(-\frac{1}{4} \right)^n \frac{(2n-1)!}{(n-1)!}\cdot
x^{-\frac{2n+1}{2}},
\end{gather}
and
\begin{equation}
  \label{eq:5}
  g_1^{(n)}(x)=\sum_{k=\lceil \frac{n}{2} \rceil}^n \frac{ (-1)^{k+1} r^k n! }{
    (1+rx^2)^k  k} \binom{k}{n-k} (2x)^{2k-n},
\end{equation}
where $\lceil n/2 \rceil$ means the smallest integer greater than or equal to
$n/2$.
The $n$th derivative of the integrand can now be determined using the
Fa\`{a} di Bruno  formula (cf.~\cite{bruno} and \cite{aldro}):
\begin{multline}
  \label{eq:6}
  I^{(n)}(x)= \sqrt{r}\cdot\sum_{k=1}^{n} h^{(k)}(g(x)) \cdot\\
\B_{nk}(g^{(1)}(x),g^{(2)}(x),\dots,g^{(n-k+1)}(x)).
\end{multline}
The {\em Bell matrices} $\B_{nk}$  used in this formula are defined by
\begin{multline}
  \label{eq:7}
  \B_{nk}(z_1,z_2,\dots,z_{n-k+1})= \\
  \sum_{\{\nu_i \} } \frac{ n! }{
    \prod_{j=1}^n [\nu_j!(j!)^{\nu_j}]} z_1^{\nu_1}z_2^{\nu_2}\dots
  z_{n-k+1}^{\nu_{n-k+1}}, 
\end{multline}
where the sum is taken over all those  sets $\{\nu_i \}$ of
non-negative integers which satisfy
\begin{equation}
  \label{eq:8}
  \sum_{j=1}^n j \nu_j=n \quad\text{and}\quad \sum_{j=1}^n \nu_j = k.
\end{equation}
To proceed we again define two
auxiliary functions
\begin{equation}
  g_3(x)=\frac{c_1}{x}\quad\text{and}\quad
  g_4(x)= \frac{c_2}{\sqrt{1-x}},
\end{equation}
such that
\begin{equation}
  \label{eq:9}
  R(f)=I(f)-g_3(f)-g_4(f).
\end{equation}
With the derivatives
\begin{gather}
  \label{eq:10}
  g_3^{(n)}(x)= c_1 \, (-1)^n\, n! \, x^{-(n+1)}\quad\text{and}\\
  g_4^{(n)}(x)= 8 c_2 \cdot \left(\frac{1}{4} \right)^{n+1}
  \frac{(2n-1)!}{(n-1)!}\cdot (1-x)^{-\frac{2n+1}{2}}
\end{gather}
we can calculate the $n$th derivative of $R(f)$ as
\begin{equation}
  \label{eq:11}
  R^{(n)}(f)=I^{(n)}(f)-g_3^{(n)}(f)- g_4^{(n)}(f).
\end{equation}

The final result then reads
\begin{multline}
\label{eq:12}
\pm\sqrt{2} x(f) = c_1 \ln f - 2 c_2 \sqrt{1-f}   \\
+\sum_{n=0}^\infty R^{(n+1)}\left(\frac{1}{2}\right)
\frac{(f-\frac{1}{2})^{n+1}-(\frac{1}{2})^{n+1}}{(n+1)!}
\;.
\end{multline}

Although for practical purposes it is much easier to calculate higher
derivatives of $R(f)$ by some computer algebra system it is
nevertheless interesting that they can indeed be given
explicitly. Convergence issues of the respective Taylor series have
not been considered here.


\bibliography{paper_master}

\end{document}